\documentclass[12pt]{article}

\usepackage{amsfonts, amsmath, amssymb, amsthm}
\usepackage{a4}
\usepackage{graphicx}
\usepackage{caption}

\DeclareMathOperator{\cn}{cn}
\DeclareMathOperator{\sn}{sn}
\DeclareMathOperator{\dn}{dn}

\author{I.G. Korepanov}
\title{Novel solutions to the tetrahedron equation}
\date{March 1989}

\begin{document}

\jot 1.5ex

\maketitle

\begin{abstract}
This is the English translation\footnote{With only this abstract and \emph{all} the footnotes added in 2013.} of the short note\footnote{I.G. Korepanov. Novel solutions to the tetrahedron equation. Chelyabinsk: Chelyabinsk Polytechnical Institute, 1989. Deposited at VINITI (\!  http://viniti.ru \!) scientific database, no.~1751-V89 (Russian).} where the first nontrivial tetrahedron relation (solution of the Zamolodchikov tetrahedron equation) with variables on the edges was presented\footnote{And the only nontrivial tetrahedron relation known by that time was due to Zamolodchikov himself.}.
\end{abstract}

\section{Formulation of the results}

Let $V_1$, $V_2$, $V_3$, $V_4$ be two-dimensional complex linear spaces with fixed bases, and $\lambda_1$, $\lambda_2$, $\lambda_3$, $\lambda_4$ be complex numbers. Consider operators
\begin{gather*}
R_{ij}^0 (\lambda_i, \lambda_j)=\begin{pmatrix} a & & & d \\  & b & c &  \\  & c & b &  \\ d & & & a \end{pmatrix},
\\
R_{ij}^1 (\lambda_i, \lambda_j)=\begin{pmatrix} -a' & & & d' \\  & -b' & c' &  \\  & -c' & b' &  \\ -d' & & & a' \end{pmatrix}
\end{gather*}
acting in $V_i \otimes V_j$, \ $1\le i<j\le 4$. Here
\begin{gather*}
a=\cn (\lambda_i-\lambda_j), \quad b=\sn (\lambda_i-\lambda_j) \dn (\lambda_i-\lambda_j),
\\
c=\dn (\lambda_i-\lambda_j), \quad d=k \sn (\lambda_i-\lambda_j) \cn (\lambda_i-\lambda_j),
\\
a'=\cn (\lambda_i+\lambda_j), \quad b'=\sn (\lambda_i+\lambda_j) \dn (\lambda_i+\lambda_j),
\\
c'=\dn (\lambda_i+\lambda_j), \quad d'=k \sn (\lambda_i+\lambda_j) \cn (\lambda_i+\lambda_j);
\end{gather*}
all elliptic functionas are of modulus~$k$.

As we show in Section~\ref{s:2}, there exist $8\times 8$ matrices $S_{123}$, $S_{124}$, $S_{134}$, $S_{234}$, whose components will be denoted as, for instance, $(S_{123})_{def}^{abc}$, \ $a,\ldots,f=0$ or~$1$, such that the relations of tetrahedral Zamolodchikov algebra~\cite{1} hold:
\begin{equation}\label{1}
R_{12}^a R_{13}^b R_{23}^c = \sum_{d,e,f=0}^1 (S_{123})_{def}^{abc} R_{23}^f R_{13}^e R_{12}^d,
\end{equation}
and similarly for the other $S$-matrices. Arguments~$\lambda_i$ are omitted in formula~\eqref{1}; it is understood also that $S_{123}$ depends on $\lambda_1,\lambda_2$ and~$\lambda_3$, and so on.

In contrast to the special case $k=0$ of paper~\cite{1}, the $S$-matrices are determined uniquely in the general case (making $k$ tend to zero, we can get unique matrices for $k=0$ as well). Introduce two-dimensional spaces $E_{12}$, $E_{13}$, $E_{14}$, $E_{23}$, $E_{24}$ and~$E_{34}$, and consider $S_{123}$ as an operator in $E_{12}\otimes E_{13}\otimes E_{23}$, \ \ $S_{124}$ as an operator in $E_{12}\otimes E_{14}\otimes E_{24}$, and so on. Does the tetrahedron equation
\[
S_{123} S_{124} S_{134} S_{234} = S_{234} S_{134} S_{124} S_{123} 
\]
hold?

The author has performed direct calculations\footnote{Using Fortran, integer arithmetic, and a Soviet-made computer Iskra. The point is that all the check, for the $S$-matrices whose elements are written out below, can be reduced to checking the equalities between polynomials with integer coefficients. And these are equal if their values coincide for many enough values of their arguments, which, again, can be taken integer.} for $k=0$, and they gave the positive answer. Below we write out matrix~$S_{123}$ explicitly, in terms of values $\tau_i=\tan \lambda_i$ (the rest of $S$-matrices are obtained by the obvious changes of subscripts). We introduce the function
\[
f(\rho,\sigma) = \frac{1+\rho\sigma}{\rho+\sigma}.
\]
So, all elements of matrix~$S_{123}$ are zeros except the following:
\jot 1.9ex
\begin{gather*}
S_{000}^{000}=S_{011}^{011}=S_{101}^{101}=S_{110}^{110}=1,
\\
S_{010}^{001}=f(\tau_1,\tau_3)f(\tau_2^{-1},\tau_3),
\\
S_{100}^{001}=f(\tau_1,-\tau_2^{-1})f(\tau_2^{-1},\tau_3),
\\
S_{111}^{001}=f(\tau_1,-\tau_2^{-1})f(\tau_1,\tau_3),
\\
S_{001}^{010}=f(\tau_2^{-1},-\tau_3)f(\tau_1,-\tau_3),
\\
S_{100}^{010}=f(\tau_1,-\tau_2^{-1})f(\tau_1,-\tau_3),
\\
S_{111}^{010}=f(\tau_1,-\tau_2^{-1})f(\tau_2^{-1},-\tau_3),
\\
S_{001}^{100}=-f(\tau_2^{-1},-\tau_3)f(\tau_1,\tau_2^{-1}),
\\
S_{010}^{100}=f(\tau_1,\tau_3)f(\tau_1,\tau_2^{-1}),
\\
S_{111}^{100}=-f(\tau_1,\tau_3)f(\tau_2^{-1},-\tau_3),
\\
S_{001}^{111}=f(\tau_1,\tau_2^{-1})f(\tau_1,-\tau_3),
\\
S_{010}^{111}=-f(\tau_1,\tau_2^{-1})f(\tau_2^{-1},\tau_3),
\\
S_{100}^{111}=-f(\tau_1,-\tau_3)f(\tau_2^{-1},\tau_3).
\end{gather*}

\begin{figure}[p]
\centering
\fboxsep 0pt
\fboxrule 0.5pt
\framebox{
\includegraphics[scale=0.6]{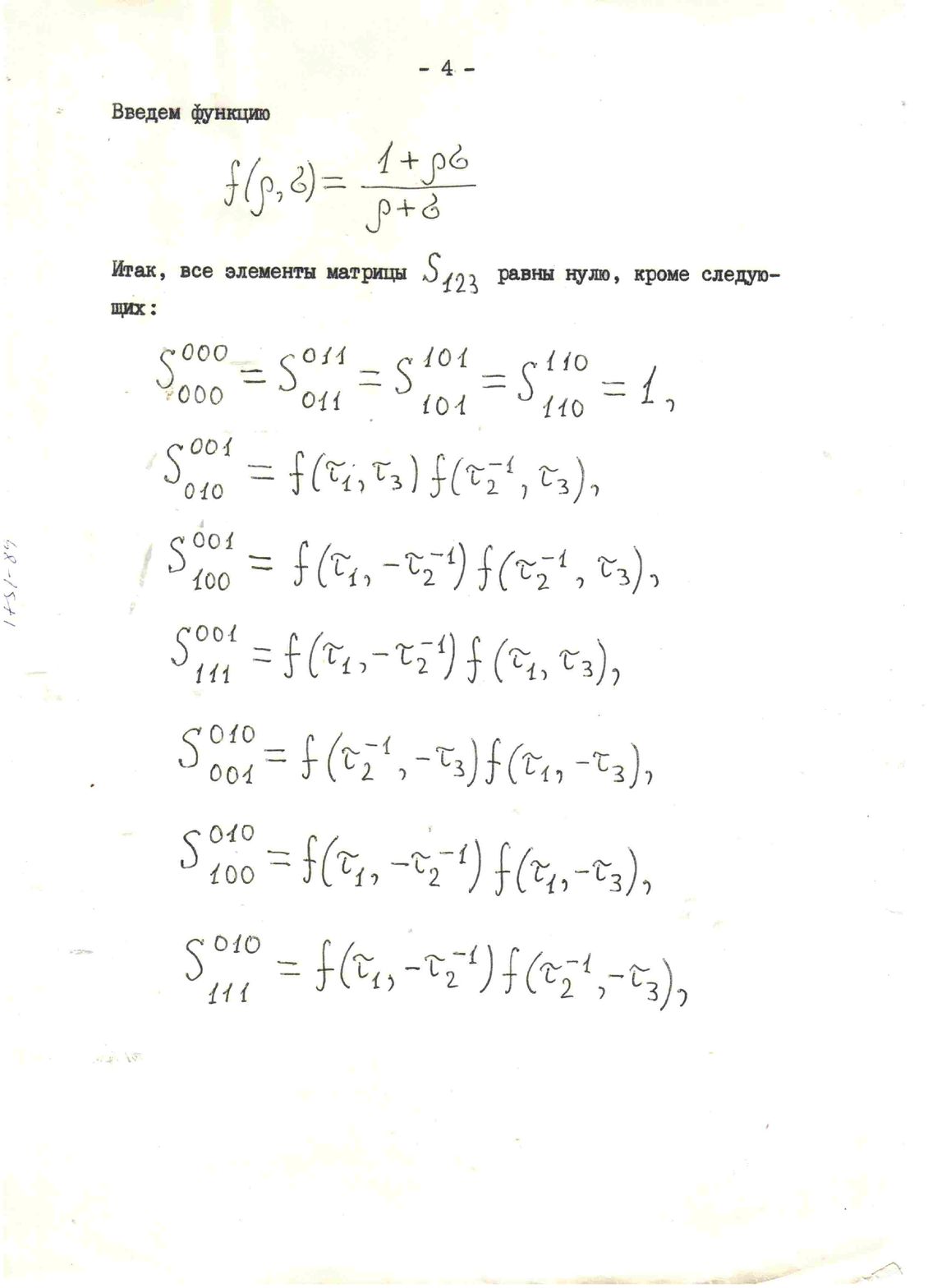}
}
\caption*{A page from the original manuscript}
\end{figure}

Recall that a model on two two-dimensional layers, related to the $S$-matrices constructed here, and whose ``Boltzmann weights'' can be made positive, has been constructed in~\cite{1}.

\section{Possible generalizations}\label{s:2}

Consider a Yang--Baxter equation
\[
R_{12}L_{01}M_{02}=M_{02}L_{01}R_{12},
\]
where the subscripts show the numbers of two-dimensional spaces in whose tensor product a given operator acts. Let $L_{01}$ and~$M_{02}$ be $L$-matrices of Felderhof~\cite{2} type. In particular, they are symmetric with respect to the transposition~${}^{\mathrm T}$. Then, as is known (see, for example,~\cite{3}), there exists, besides a symmetric matrix $R_{12}=R_{12}^0$, also a non-symmetric~$R_{12}^1$, for which
\[
(R_{12}^1)^{\mathrm T}L_{01}M_{02}=M_{02}L_{01}R_{12}^1.
\]

It can be shown, by developing ideas of paper~\cite{3}, that the space of operators~$\mathcal R_{123}$ performing the following permutation of $L$-matrices:
\[
(\mathcal R_{123})^{\mathrm T} L_{01}M_{02}N_{03} = N_{03}M_{02}L_{01} \mathcal R_{123},
\]
is eight-dimensional. The author hopes to explain this in more detail in another paper. Exactly 8 such operators~$\mathcal R_{123}$ are obtained, on one hand, in the form
\[
R_{12}^a R_{13}^b R_{23}^c,
\]
and on another hand --- in the form
\[
R_{23}^f R_{13}^e R_{12}^d,
\]
and this is what leads to linear dependencies of type~\eqref{1}. Checking the validity of the tetrahedron equation for thus obtained $S$-matrices is, however, extremely difficult and has not been done as yet.


\begin{thebibliography}{99}

\bibitem{1}
Korepanov I.G. {\it Tetrahedral analogue of Zamolodchikov algebra and a two-layer model of two-dimensional statistical physics}. Chelyabinsk Polytechnical Institute, Chelyabinsk, 1988, 9~p., Deposited at VINITI 06 June 1988, no.~4433-V88 (Russian).

\bibitem{2}
Felderhof B.U. {\it Diagonalization of the transfer matrix of the free fermion model}. Physica, 1973, vol.~66, no.~2, pp. 279--298.

\bibitem{3}
Krichever I.M. {\it Baxter equations and algebraic geometry}. Funct. Anal. and Apps., 1981, vol.~15, issue~2, pp. 92--103.

\end{thebibliography}
\end{document}